\documentclass[prb,twocolumn,showpacs]{revtex4}
\usepackage{graphicx}
\usepackage{amssymb}
\usepackage{amsmath}
\usepackage{xspace}
\usepackage{dcolumn}
\usepackage{bm}

\newfont{\tensy}{cmsy10}
\newcommand{\chem}[1]{{$\fontdimen16\tensy=3.0pt
    \fontdimen17\tensy=3.0pt \mathrm{#1}$}}
\newcommand{\ie}[0]{i.e.\@\xspace}
\newcommand{\eg}[0]{e.g.\@\xspace}
\newcommand{\omb}[0]{\bar{\omega}}
\newcommand{\etal}[0]{et al.\@\xspace}
\newcommand{\om}[0]{\omega}
\newcommand{\si}[0]{\sigma}
\newcommand{\en}[0]{\epsilon}
\renewcommand{\sf}[0]{A(k,\en)}
\newcommand{\sfb}[0]{A(\bm{k},\en)}
\newcommand{\las}[0]{\langle}
\newcommand{\ras}[0]{\rangle}
\newcommand{\la}[0]{\left\las}
\newcommand{\ra}[0]{\right\ras}
\newcommand{\ket}[1]{\left|#1\ra}  
\newcommand{\bra}[1]{\la#1\right|} 
\newcommand{\sket}[1]{|#1\ras}  
\newcommand{\sbra}[1]{\las#1|} 
\newcommand{\rme}{\mathrm{e}}

\newcommand{\rmi}{\mathrm{i}}
\newcommand{\up}{\uparrow}

\renewcommand{\Im}[0]{\text{Im}}
\newcommand{\bk}[0]{\bm{k}}
\newcommand{\ek}[0]{\en_{\bk}}
\newcommand{\G}[0]{\mathcal{G}}

\begin{document}


\title{Spectral function of electron-phonon models by cluster perturbation theory}

\author{Martin Hohenadler}\author{Markus Aichhorn}
\author{Wolfgang \surname{von der Linden}}
\affiliation{%
  Institute for Theoretical Physics, Graz University of Technology,
  Petersgasse 16, A-8010 Graz, Austria} \email{hohenadler@itp.tu-graz.ac.at}
\begin{abstract}
  Cluster perturbation theory in combination with the Lanczos method is used
  to compute the one-electron spectral function of the Holstein polaron in
  one and two dimensions. It is shown that the method allows reliable
  calculations using relatively small clusters, and at the same time
  significantly reduces finite-size effects. Results are compared with exact
  data and the relation to existing work is discussed. We also use a
  strong-coupling perturbation theory--equivalent to the Hubbard I
  approximation--to calculate the spectral function of the quarter-filled
  Holstein model of spinless fermions, starting from the exact atomic-limit
  Green function. The results agree well with previous calculations within
  the many-body coherent potential approximation.
\end{abstract}


\maketitle

\section{\label{sec:introduction}Introduction}

Spectral properties, such as the one-electron spectral function, provide
valuable insight into the usually complex physics of strongly correlated
systems. However, reliable results for such quantities are difficult to
obtain. Analytical methods are often restricted to very simple limiting
cases, and results can usually not be extended to more general situations.
Two remarkable exceptions are the Holstein model with linear electron
dispersion\cite{GuMeSc94,MeScGu94} and the Hubbard model,\cite{LiWu68} both
in one dimension, which have been solved exactly.  To study more general
models, numerical methods such as Exact Diagonalization (ED) and Quantum
Monte Carlo (QMC) have received much attention over the last decades. ED
methods allow very accurate calculations of ground-state as well as
finite-temperature spectral properties, but are restricted to rather small
clusters due to the large dimension of the corresponding Hilbert space. QMC
can be used to obtain results on large clusters even in higher dimensions,
but here the sign-problem and the ill-posed analytic continuation required to
obtain dynamic correlation functions for real frequencies are detrimental for
many interesting applications.

The recently developed cluster perturbation theory\cite{SePePL00,SePePl02}
(CPT) marks an important improvement of the situation. It is based on a
break-up of the infinite system into identical, finite clusters, on which the
one-electron Green function is calculated exactly. Then, the hopping between
clusters is treated within strong-coupling perturbation
theory\cite{PaSeTr98,PaSeTr00} (SCPT). The method has been successfully used
for various Hubbard models, to calculate spectral functions, as well as other
quantities of interest, both for zero\cite{SePePL00,SePePl02,DaArHa02} and
finite temperature.\cite{AiDaEvvdL02} Although the concept of CPT relies on a
model with local interactions only, it has also been applied with some
success to the $t-J$ model.\cite{ZaEdArHa00,ZaEdArHa02} For the calculation
of the cluster Green function, the ED Lanczos method (for a review see
Ref.~\onlinecite{Da94}) can be used.

For the case of coupled electron-phonon systems, such as the Holstein model
and its various extensions--\eg the Holstein-Hubbard or the Holstein
double-exchange model-- the application of ED methods is hampered by the
infinite number of possible phonon configurations, which gives rise to a
rapidly growing requirement of computer memory and/or CPU time as the number
of lattice sites or phonon states increases.  Consequently, standard ED
methods--employing some kind of Hilbert space truncation--are restricted to
very small clusters, especially for low phonon frequency and/or strong
electron-phonon coupling. Again improved methods such as DMRG, or the use of
variational phonon bases allow to extend the accessible parameter range.
Nevertheless, as electron-phonon interaction has been identified as an
important ingredient in, \eg, high-temperature
superconductors\cite{BYMdLBi92} and manganites\cite{David_AiP}, further
progress along these lines is highly desirable.

In this paper, we show that CPT can be successfully applied to
electron-phonon models with a (local) coupling of the Holstein
type.\cite{Ho59a} We present results for the one-electron spectral function
of the Holstein polaron, \ie the Holstein model with one electron, in one and
two dimensions. The Holstein polaron problem has been investigated
intensively in the past, and a wealth of information about its spectral
properties is available. We find that the use of CPT strongly reduces
finite-size effects, giving results which are much closer to the
thermodynamic limit than the corresponding ED data.  Additionally, we
consider the special case of a completely saturated ferromagnetic state at
zero temperature in the Holstein double-exchange model\cite{Gr01} for
colossal magnetoresistive manganites.  The latter is then equivalent to the
Holstein model of spinless fermions, and we combine the exact atomic-limit
one-particle Green function with CPT for a single-site cluster to calculate
the spectral function at quarter-filling.  The results of this simple
approach agree well with the previously developed many-body coherent
potential approximation.\cite{EdGrKu99,GrEd99,Gr01,HoEd01,David_AiP}

The paper is organized as follows: In Sec.~\ref{sec:cpt} we give a review of
CPT. Sec.~\ref{sec:holsteinpolaron} discusses the application to the Holstein
polaron, while the SCPT for the many-electron case is presented in
Sec.~\ref{sec:holsteinDE}. Finally, Sec.~\ref{sec:summary} contains our
conclusions.

\section{\label{sec:cpt}Cluster perturbation theory}

The basic idea of CPT\cite{SePePL00,SePePl02} is to divide the infinite
lattice into identical clusters, each containing $N$ lattice sites.  Adopting
the notation of Ref.~\onlinecite{SePePl02}, the Hamiltonian of the system is
written in the form $H = H_0 + V$ where
\begin{equation}\label{eq:H0pV}
H_0 = \sum_{\bm{R}} H^{\bm{R}}_0\,,\quad
V   = \sum_{\overset{\bm{R},\bm{R'}}{a,b}} V^{\bm{R},\bm{R'}}_{a,b}
c^\dag_{\bm{R}a}c^{\phantom\dag}_{\bm{R'}b}\,.
\end{equation}
In the most general formulation of CPT,\cite{SePePl02} the subscripts $a$,
$b$ denote different orbitals within a cluster, but here we restrict
ourselves to the case of one orbital per site so that $a$, $b=1,\dots,N$. The
vectors $\bm{R}$, $\bm{R'}$ correspond to sites in the superlattice of
clusters (see Ref.~\onlinecite{SePePl02}). In Eq.~(\ref{eq:H0pV}),
$H^{\bm{R}}_0$ represents a Hamiltonian of a single cluster--containing local
interactions only--and $V$ describes the hopping between clusters, \ie the
hopping amplitude between site $a$ of cluster $\bm{R}$ and site $b$ of
cluster $\bm{R'}$ is given by the matrix element $V^{\bm{R},\bm{R'}}_{a,b}$.
Although long-range hopping can also be included,\cite{SePePl02} we shall
only consider models with nearest-neighbor hopping so that
$V^{\bm{R},\bm{R'}}_{a,b}=-t$ for neighboring sites $a$, $b$ in adjacent
clusters $\bm{R}$, $\bm{R'}$.  Within CPT, an approximation for the Green
function of the original system, $\G(\bk,\en)$, is obtained using an
analytical strong-coupling perturbation expansion up to first order of the
inter-cluster hopping $V$ (for details of the derivation see
Ref.~\onlinecite{SePePl02}). The resulting equation relating the Green
function of the original lattice to the energy-dependent cluster Green
function $G(z)$ reads\cite{SePePL00,SePePl02}
\begin{equation}\label{eq:CPT1}
  \G_{ab}(\bm{Q},z) = \left(
    \frac{G(z)}{1-V(\bm{Q})G(z)}
  \right)_{a,b}\,.
\end{equation}
Here $z=\en + \rmi\eta$, $G(z)$ and $V(\bm{Q})$ stand for $N\times N$
matrices, and the inter-cluster hopping $V$ has been partially
Fourier-transformed exploiting the translational symmetry of the cluster
superlattice, with $\bm{Q}$ being a wave vector of the reduced Brillouin
zone.\cite{SePePl02} Finally, the Green function $\G_{ab}$ can be transformed
from the mixed representation of Eq.~(\ref{eq:CPT1}), real space within a
cluster and reciprocal space between clusters, using\cite{SePePl02}
\begin{equation}\label{eq:CPT2}
\G(\bk,z) = \frac{1}{N}\sum_{a,b=1}^N
\G_{ab}(\bk,z)\rme^{-\rmi\bk\cdot(\bm{r}_a-\bm{r}_b)}
\end{equation}
to obtain the familiar representation of the one-electron Green function.
$\G(\bk,z)$ as given by Eqs.~(\ref{eq:CPT1}) and~(\ref{eq:CPT2}) becomes
exact in the atomic limit $t=0$ (Refs.~\onlinecite{SePePL00,SePePl02}).
Moreover, it also reduces to the exact result for the case of non-interacting
electrons\cite{SePePL00,SePePl02} since, in this case, Eq.~(\ref{eq:CPT1})
corresponds to the exact resummation of the perturbation series. Finally, CPT
also becomes exact in the limit $N\rightarrow\infty$
(Refs.~\onlinecite{SePePL00,SePePl02}). The one-electron cluster Green
function at zero temperature
\begin{eqnarray}\nonumber
  G_{ab}(\en) = &&
  \bra{\Omega}c_{a}      \frac{1}{z-(H_0-E_0)} c^\dag_{b}\ket{\Omega}\\
  && +
  \bra{\Omega}c^\dag_{b} \frac{1}{z+(H_0-E_0)} c_{a}\ket{\Omega}
\end{eqnarray}
can be calculated exactly for any pair of site indices $a$, $b$ in the
cluster using, \eg, the Lanczos method. Here $E_0$ is the energy of the
ground state $\ket{\Omega}$ of the cluster, and a spin index has been
suppressed in the notation. The two parts of the Green function matrix
$G_{ab}$ correspond to adding or removing an electron to/from $\ket{\Omega}$.
Finally, the one-electron spectral function is defined as
\begin{equation}\label{eq:Akw}
\sfb = -\pi^{-1}\lim_{\eta\rightarrow 0^+}\Im\,\G(\bk,\en+\rmi\eta)\,.
\end{equation}

Since CPT is based on a perturbation expansion in the inter-cluster hopping,
the method can be expected to work especially well in the strong-coupling
regime. This is also illustrated by the fact that it becomes exact in the
atomic limit, as mentioned above. On the other hand, for weak or intermediate
coupling, the electronic kinetic energy is not small compared to the local
interactions. Consequently, the size of the cluster has to be large enough in
order to obtain accurate results. In fact, from previous applications of CPT,
\eg, to the one- and two-dimensional Hubbard
model,\cite{SePePL00,SePePl02,DaArHa02} the cluster size $N$ emerged as the
main control parameter of the method.  In the case of the one-dimensional
Hubbard model, for example, $N=1$ is identical to the Hubbard I
approximation,\cite{Hu63} while $N=2$ already gives a spectral function that
contains most of the relevant features such as short-range antiferromagnetic
ordering.\cite{SePePl02} With increasing $N$, the CPT Green function
approaches systematically the exact result for the infinite system. For
identical cluster size, the CPT spectrum contains many more poles with
significant residues than the corresponding results of ED. In fact, also in
the 1D Hubbard model, the spectrum obtained with CPT on a four-site cluster
is already comparable in quality to the ED spectrum for $N=12$
(Refs.~\onlinecite{SePePL00,SePePl02}). An additional advantage of CPT is the
possibility to evaluate $\sfb$ at continuous wavevectors $\bk$, in contrast
to ED which restricts $\bk$ to the $N$ vectors of the first Brillouin zone of
the cluster, of which only $N/2+1$ are physically distinct. Finally, finite
temperature Lanczos methods can also be combined with CPT to calculate
thermodynamic properties.\cite{AiDaEvvdL02}

Concerning the application of the Lanczos method to calculate the cluster
Green function $G_{ab}$, it is important to stress the need for open boundary
conditions (BCs). Attempts have been made to use periodic BCs and subtract
the corresponding terms afterward in the perturbative treatment of the
inter-cluster hopping, but it has been found that the accuracy of the results
is much better for the case of open BCs. Although the latter are physically
more intuitive in connection with CPT, the calculation of the cluster Green
function with the Lanczos method becomes more difficult as one cannot exploit
translational symmetry. Other symmetries such as the inversion group can be
used in principle, but are usually not as effective in saving computer memory
by reducing the size of the corresponding Hamiltonian matrix and Lanczos
vectors.

We want to point out that CPT does not in principle rely on the ED method. In
fact, the cluster Green function may be calculated using any method
available.\cite{SePePl02} Indeed we will see in Sec.~\ref{sec:holsteinDE}
that it is possible to combine the exact analytic solution for the
atomic-limit Green function with CPT, to obtain results which agree
surprisingly well with the many-body coherent potential
approximation.\cite{HoEd01}

In addition to the spectral function considered here, other physical
properties of the system can also be calculated with CPT. This includes, \eg,
the ground state energy of the infinite system, the electronic kinetic energy
or the Fermi surface.\cite{SePePl02} The strength of CPT lies in the
calculation of the one-particle Green function and related quantities such as
the density of states. The numerical effort is relatively small compared to
more sophisticated methods like DMRG or QMC. An additional advantage of CPT
is the fact that it can easily be applied also to two-dimensional systems, in
contrast to, \eg, DMRG. Finally, an important disadvantage of CPT should be
mentioned: Within the current formulation, two-particle Green functions
cannot be calculated. Consequently, it is not possible to compute, \eg, the
dc conductivity or other interesting two-particle correlation functions.

\section{\label{sec:holsteinpolaron}Holstein polaron}

The Hamiltonian of the Holstein model\cite{Ho59a} reads
\begin{eqnarray}\label{eq:hamilt}\nonumber
H =
   && -t\sum_{\las ij\ras\si} (c^\dag_{i\si}c^{\phantom{\dag}}_{j\si} + \text{h.c})
      +\om\sum_i b^\dag_i b_i\\
   && -g\sum_i n_i (b^\dag_i+b_i)\,,
\end{eqnarray}
where $c^\dag_{i\si}$ ($c^{\phantom{\dag}}_{i\si}$) and $b^\dag_i$ ($b_i$)
are creation (annihilation) operators for an electron with spin $\si$ and a
phonon of frequency $\om$ at lattice site $i$, respectively. The electron
occupation number is defined as $n_i = \sum_\si n_{i\si}$ with
$n_{i\si}=c^\dag_{i\si}c_{i\si}$, and the parameters of the model are the
hopping integral for nearest-neighbor hopping, $t$, and the electron-phonon
coupling strength, denoted as $g$. It is common to define a dimensionless
coupling parameter $\lambda=g^2/(2\om W)$, where $2W$ is the bandwidth of the
bare electron band, and a dimensionless phonon frequency $\omb=\om/t$. The
Holstein model can then be described using only these two parameters.
Moreover, we shall express all energies in units of $t$. As mentioned above,
here we only consider the one-electron limit of Hamiltonian~(\ref{eq:hamilt})
which is also called the Holstein polaron problem. Although there is only a
single electron in the system, the coupling to the phonons makes it a complex
many-body problem, which has been the focus of much theoretical work.  The
restriction to one electron greatly simplifies calculations with the Lanczos
method since both, the number of required phonon states\cite{Marsiglio95} and
the number of electron configurations grow noticeably with the number of
particles. However, in Sec.~\ref{sec:holsteinDE}, we will use the exact
result for the atomic-limit Green function and CPT for a single-site cluster
to calculate the spectral function of the Holstein model of spinless fermions
at quarter-filling.

Following other
authors,\cite{RaTh92,Marsiglio93,AlKaRa94,FeLoWe97,Robin97,dMeRa97} we
calculate the Green function
\begin{equation}\label{eq:G_lanczos}
  \G(\bk,\en) = \bra{0}c_{\bk}\frac{1}{\en-H} c^\dag_{\bk}\ket{0}\,,
\end{equation}
where $\ket{0}$ represents the ground state of the phonons and the vacuum
state for the electrons. The spin index can be suppressed due to the symmetry
of the problem. The corresponding one-electron spectral function is given by
Eq.~(\ref{eq:Akw}).

Compared to the class of Hubbard models for which CPT has been originally
developed, we are facing an additional difficulty arising from the \`{a}
priori infinite number of allowed phonon states. We employ a widely used
truncation scheme\cite{Marsiglio93} of the phonon Hilbert space which is
spanned by the basis states
\begin{equation}\label{eq:basis}
  \ket{r}_\text{ph} =
  \prod_{i=1}^N\frac{1}{\sqrt{\nu_i^{(r)}!}}
  \left(b^\dag_i\right)^{\nu_i^{(r)}}
  \ket{0}_\text{ph}\,,
\end{equation}
where $\nu^{(r)}_i$ denotes the number of phonons at lattice site $i$. Now
the truncation consists in restricting the basis states to the subset with
\begin{equation}
  \sum_{i=1}^N \nu^{(r)}_i\leq N_\text{ph}
\end{equation}
leading to $(N_\text{ph}+N-1)!/(N_\text{ph}!(N-1)!)$ allowed phonon
configurations.  The convergence of the results with $N_\text{ph}$ can be
monitored using the ground-state energy $E$ of the cluster with open BCs. In
all results of this paper, $N_\text{ph}$ was chosen such that the relative
error for the ground state with one electron,
$|E(N_\text{ph}+1)-E(N_\text{ph})|/|E(N_\text{ph})|$, was smaller than
$10^{-5}$. We find that convergence of $E$ also ensures a well-converged
spectral function. Moreover, the influence of the number of phonons kept in
the calculation is much larger for the incoherent part of the spectrum than
for the coherent, low-energy quasi-particle peak which determines $E$ (see
Sec.~\ref{sec:comp-with-exact}).  Finally, a refined truncation scheme which
allows for extremely accurate results (relative error $<10^{-7}$) has been
proposed by Wellein \etal\cite{WeRoFe96}.

Before we come to a discussion of the results obtained with CPT, we want to
comment on some of the existing work on spectral properties of the Holstein
polaron. As indicated before, the most reliable method to calculate dynamic
quantities, such as $\sfb$, is ED which has been used extensively in the
past.
\cite{RaTh92,Marsiglio93,AlKaRa94,WeRoFe96,WeFe97,FeLoWe97,FeLoWe00,dMeRa97,Robin97}
Most of this work has focused on the polaron band structure $E(\bk)$ instead
of the spectral function, since it is often easier to interpret, especially
in the strong-coupling regime where the structure of $\sfb$ is rather
complicated. However, as pointed out by Wellein \etal,\cite{WeRoFe96} the two
quantities are closely related. In fact the position of the lowest-energy
peak in $\sfb$, obtained from the Green function~(\ref{eq:G_lanczos}),
follows exactly the polaron band structure as we vary $\bk$. Moreover, as
discussed by Wellein and Fehske,\cite{WeFe97} the integral over this first
peak is equivalent to the quasi-particle (QP) weight
$z(\bk)=|\sbra{\psi^{(1)}_{0,\bk}}c^\dag_{\bk}\sket{0}|^2$, where
$\psi^{(1)}_{0,\bk}$ denotes the lowest-energy single-polaron state in the
sector with total momentum $\bk$. Other numerical methods which have been
used to calculate the spectrum of the Holstein polaron include
DMRG\cite{JeWh98,ZhJeWh99} (in one dimension), finite-cluster strong coupling
perturbation theory\cite{Stephan} (1D, 2D), QMC\cite{Ko98,Ko99} (1D--3D), and
variational methods\cite{RoBrLi99I,BoTrBa99,KuTrBo02} (1D--4D).

\subsection{Comparison with Exact Diagonalization}\label{sec:comp-with-exact}

\begin{figure}
  \includegraphics[width=0.45\textwidth]{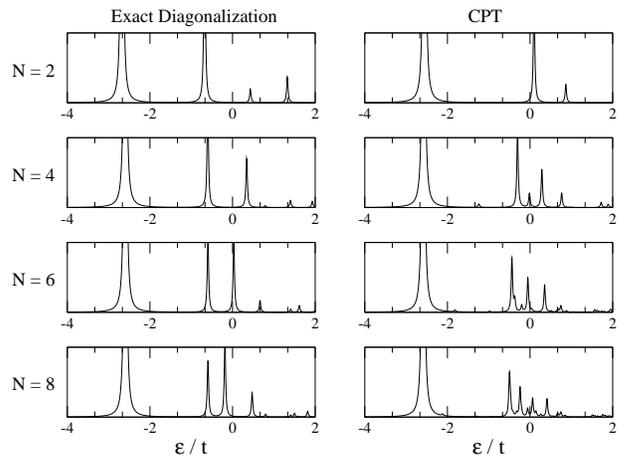}
\caption{\label{fig:lan_cpt_weak}
  Comparison of the spectral function $A(0,\en)$ of the 1D Holstein polaron
  obtained with ED (left column) and CPT (right column) for various numbers
  of lattice sites $N$ in the cluster. The plot is for $\omb=2.0$,
  $\lambda=0.5$ and $N_\text{ph}=6$. An artificial imaginary part
  $\eta=0.02t$ has been used to broaden the delta peaks.}
\end{figure}

As mentioned above, the critical parameter of CPT is the number of sites in
the cluster. To demonstrate the advantage of CPT over the standard ED method
(see, \eg, Ref.~\onlinecite{Marsiglio93}) we present in
Fig.~\ref{fig:lan_cpt_weak} the spectral function $A(0,\en)$ in one dimension
for different cluster sizes $N$. We chose $\omb=2$ and $\lambda=0.5$, which
is the regime where an extended polaron exists (see, \eg,
Ref.~\onlinecite{WeRoFe96}).  Consequently, significant finite-size effects
can be expected for small clusters, which is exactly what we see in the ED
results. For the latter periodic BCs have been used.
Fig.~\ref{fig:lan_cpt_weak} clearly shows that the shape of the large QP peak
at $\en\approx-2.4$ changes very little with increasing $N$ for both, ED and
CPT, but a noticeable shift can be observed in the case of the ED spectra as
we go from $N=2$ to $N=4$. The influence of $N$ is much larger for the
incoherent part of the spectrum, which lies about a distance $\omb$ above the
QP peak.  The ED spectra display sharp, well-separated peaks, whereas the
corresponding CPT data--containing many more poles--resembles much closer the
expected results for an infinite system. The latter has been investigated by
Marsiglio\cite{Marsiglio93} using Migdal-Eliashberg theory. For the same
parameters, he found that the QP peak remains almost unchanged as
$N\rightarrow\infty$, while the incoherent part evolves into a continuous
band that fits well to the CPT results even for rather small clusters
$N\gtrsim6$. We have also compared $\sf$ for $k\neq0$, and the observed
influence of finite-size effects agrees perfectly with previous work of
Wellein and Fehske:\cite{WeFe97} As $k$ increases from $k=0$ to $k=\pi$, the
size of the polaron increases, and the deviations of the ED data from the CPT
results become larger. In the strong-coupling or small-polaron regime, not
shown here, finite-size effects are known to be small. Consequently, even for
very small clusters, ED and CPT both give well-converged results for the QP
peak which determines, \eg, the ground-state energy.  However, in the case of
ED, the incoherent part of the spectrum for wavevector $k$, corresponding to
excitations of an electron with momentum $q$ and a phonon with momentum
$k-q$, still exhibits the typical multi-peak structure of a finite system,
whereas the CPT results again reproduce much better the incoherent band found
in the thermodynamic limit. Moreover, as mentioned in Sec.~\ref{sec:cpt}, CPT
allows to calculate $\sf$ for continuous $k$, while ED on a $N$-site cluster
is restricted to $N/2+1$ physically non-equivalent wavevectors.

A closer look at the CPT results in Fig.~\ref{fig:lan_cpt_weak} reveals small
additional peaks--not present in the ED spectra--which move from the
incoherent part of $\sf$ towards the QP peak with increasing $N$.  Additional
calculations for larger clusters have shown that these peaks vanish
systematically with increasing $N$, so that the CPT spectrum approaches the
exact result in the thermodynamic limit $N=\infty$, as expected.
Consequently, these peaks are not a defect of CPT, but represent finite-size
effects which arise from the approximate treatment of inter-cluster hopping.
The latter, in combination with the open BCs used to calculate the cluster
Green function, leads to a system which does not have perfect translational
symmetry. The situation is equivalent to ED with open BCs: For
$N\rightarrow\infty$ the spectrum approaches the results of an infinite
cluster. However, in contrast to CPT, the effects for finite $N$ are much
more significant. Moreover, these finite-size effects manifest themselves in
a slightly different way than in the case of periodic BCs, where no
additional peaks--showing the afore mentioned behavior--are found.  In the
case of CPT, already for the small cluster sizes shown in
Fig.~\ref{fig:lan_cpt_weak}, the spectral weight of these additional peaks is
extremely small compared to the rest of the spectrum. For other values of
$\omb$ and $\lambda$, a similar behavior has been found. Although not
discussed by the authors, similar effects can also be expected for the case
of the Hubbard model,\cite{SePePL00,SePePl02,DaArHa02} although they may be
larger for the Holstein polaron due to the higher sensitivity of
phonon-excitations to the BCs.

\subsection{Results: One dimension}\label{sec:1d-holstein-polaron}

\begin{figure}
  \includegraphics[width=0.41\textwidth]{1d_0.8_0.25_sf.eps}\\\vspace*{1em}
  \includegraphics[width=0.45\textwidth,height=0.3\textwidth]{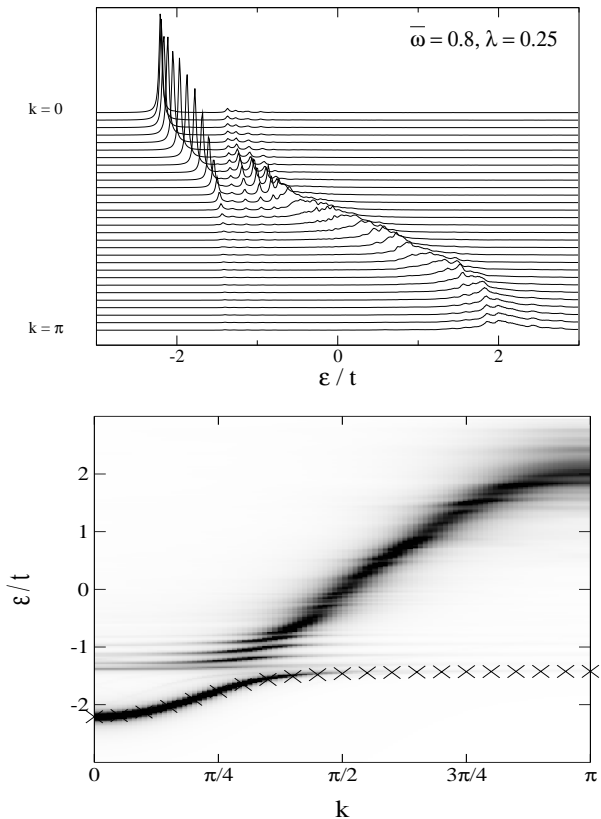}
\caption{\label{fig:spectrum1D_lambda0.25}
  Top: Spectral function $\sf$ of the 1D Holstein polaron calculated with CPT
  for $N=14$, $N_\text{ph}=6$ and $\eta=0.02t$. Bottom: Density plot of the
  same data for 100 points in $k$ space. Symbols represent results of
  Bon\v{c}a \etal\cite{BoTrBa99}}
\end{figure}

In one dimension, the general picture emerging from previous work on the
Holstein polaron problem is as follows (see \eg Ref.~\onlinecite{WeRoFe96}
and references therein): In the non-adiabatic regime ($\omb>1$), a so-called
Lang-Firsov polaron is formed which, due to the instant response of the
phonons to the electronic motion, represents a very localized object.  As the
electron-phonon coupling increases, its mobility or effective hopping
amplitude exhibits a gradual decrease, and for strong coupling a
nearly-localized small polaron moving in an exponentially narrow band exists.
In contrast, in the adiabatic ($\omb<1$), weak-coupling regime the electron
drags with it an extended cloud of phonons. This object is usually called a
`large polaron'. At $\lambda\approx1$, a sharp but continuous\cite{Loe88}
transition to a less mobile small Holstein polaron takes place. The two
conditions for small-polaron formation, independent of the value of $\omb$,
are $\lambda>1$ {\it and} $\lambda/\omb>0.5$.\cite{WeRoFe96} Moreover, as
pointed out by Capone \etal,\cite{CaStGr97} the formation of small polarons
is determined by $\lambda>1$ for $\omb<1$ and by $\lambda/\omb>0.5$ for
$\omb>1$.  Here we restrict ourselves to the most interesting regime of
phonon energies comparable to the electronic hopping, \ie $\omb\sim 1$. For
intermediate $\omb$ and $\lambda$, no reliable analytical methods exist, so
that numerical approaches represent the most important source of information.

In Fig.~\ref{fig:spectrum1D_lambda0.25} we present results for $\sf$ for
$\omb=0.8$, $\lambda=0.25$ and $N=14$, as well as a density plot of the same
data. As mentioned before, the spectrum consists of a low-lying QP peak and
an incoherent part at higher energies. The physics behind the observed
behavior of $\sf$ has been discussed, \eg, by Stephan,\cite{Stephan} and is
typical for electronic systems weakly interacting with dispersionless optical
phonons. For small $k$, most of the spectral weight resides in the QP peak
which corresponds to a weakly-dressed electron. For the case considered here,
in which the phonon energy lies inside the bare electron band, electron and
phonon hybridize and repel each other near the point where they would be
degenerate, \ie for $|E(k)-E(0)|\sim\omb$. This coincides with the region
where the flattening of the polaron band occurs and, in fact, for larger $k$
the phonon becomes the lowest-energy excitation. However, most of the
spectral weight is contained in the broad, incoherent band which follows the
free-electron dispersion. The density plot in
Fig.~\ref{fig:spectrum1D_lambda0.25} also contains data for the polaron band
structure $E(k)$ which has been obtained by Bon\v{c}a \etal\cite{BoTrBa99}
using their variational ED method. The latter has been shown to give very
accurate results for the infinite system, although it becomes somewhat less
accurate in the strong coupling regime and for large values of $k$
(Ref.~\onlinecite{BoTrBa99}). As mentioned before, $E(k)$ corresponds to the
lowest-energy band in $\sf$ and we find very good agreement with our data
throughout the Brillouin zone.

\begin{figure}
  \includegraphics[width=0.41\textwidth]{1d_1.0_0.5_sf.eps}\\\vspace*{1em}
  \includegraphics[width=0.45\textwidth,height=0.3\textwidth]{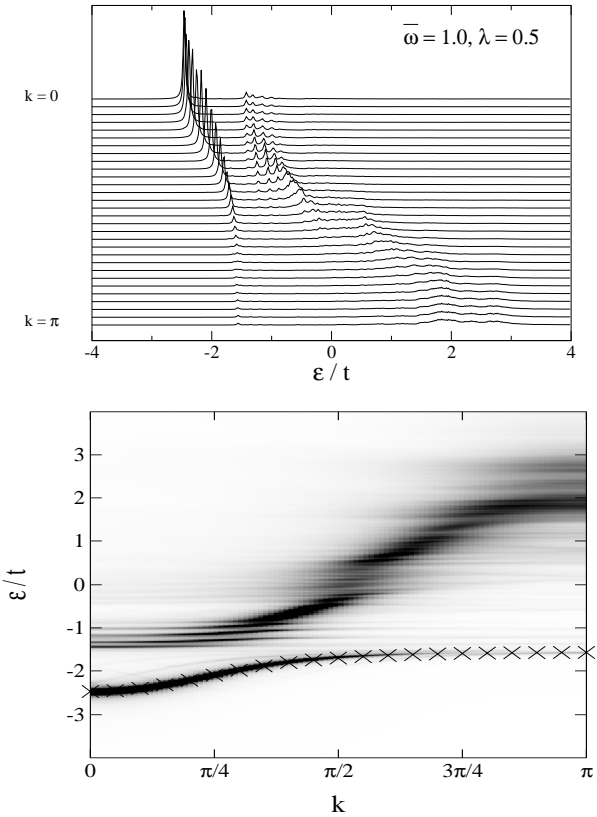}
\caption{\label{fig:spectrum1D_lambda0.5}
  Top: Spectral function $\sf$ of the 1D Holstein polaron calculated with CPT
  for $N=12$, $N_\text{ph}=6$ and $\eta=0.02t$. Bottom: Density plot of the
  same data for 100 points in $k$ space. Symbols represent results of
  Bon\v{c}a \etal\cite{BoTrBa99}}
\end{figure}

Fig.~\ref{fig:spectrum1D_lambda0.5} shows results for a similar phonon
frequency $\omb=1.0$ but for stronger electron-phonon coupling $\lambda=0.5$
and $N=12$. Compared to the weak-coupling case discussed above, the polaron
band is separated more clearly from the incoherent part of the spectrum and,
as expected, the band-width is further reduced.  Additionally, even more
spectral weight has been transfered to the high-energy, incoherent band.  On
top of that, a gap shows up in the upper band at about $k=\pi/2$.  Again the
polaron band fits very well the results for $E(k)$ of Bon\v{c}a
\etal\cite{BoTrBa99}

\begin{figure}
  \includegraphics[width=0.41\textwidth]{1d_1.0_1.0_sf.eps}\\\vspace*{1em}
  \includegraphics[width=0.45\textwidth,height=0.3\textwidth]{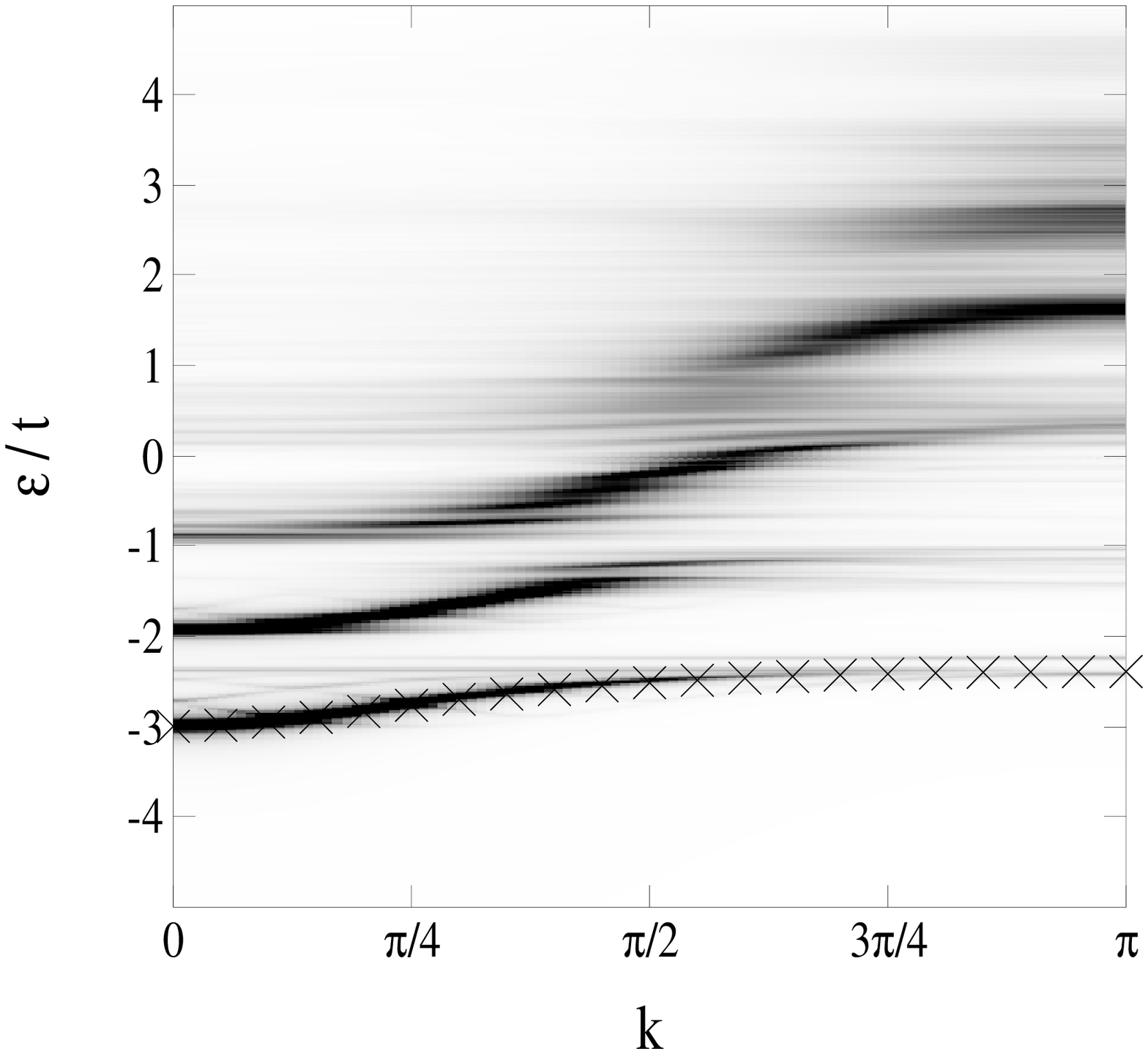}
\caption{\label{fig:spectrum1D_lambda1.0}
  Top: Spectral function $\sf$ of the 1D Holstein polaron calculated with CPT
  for $N=8$, $N_\text{ph}=9$ and $\eta=0.02t$. Bottom: Density plot of the
  same data for $100$ points in $k$ space. Symbols represent results of
  Bon\v{c}a.\cite{Bo03}}
\end{figure}

We next consider the case of intermediate coupling $\lambda=1.0$, with
$\omb=1$ and $N=8$ (Fig.~\ref{fig:spectrum1D_lambda1.0}). For these
parameters, an extended polaron exists which still has a relatively large
band-width, compared to the small-polaron case discussed below.  Moreover,
the incoherent part of the spectrum has split up into several sub-bands
separated in energy by $\omb$, which correspond to excitations of an electron
and one or more phonons. As before, we find very good agreement between the
low-energy band in $\sf$ and the polaron band energy $E(k)$ calculated by
Bon\v{c}a.\cite{Bo03}

\begin{figure}
  \includegraphics[width=0.41\textwidth]{1d_1.0_2.0_sf.eps}\\\vspace*{1em}
  \includegraphics[width=0.45\textwidth,height=0.3\textwidth]{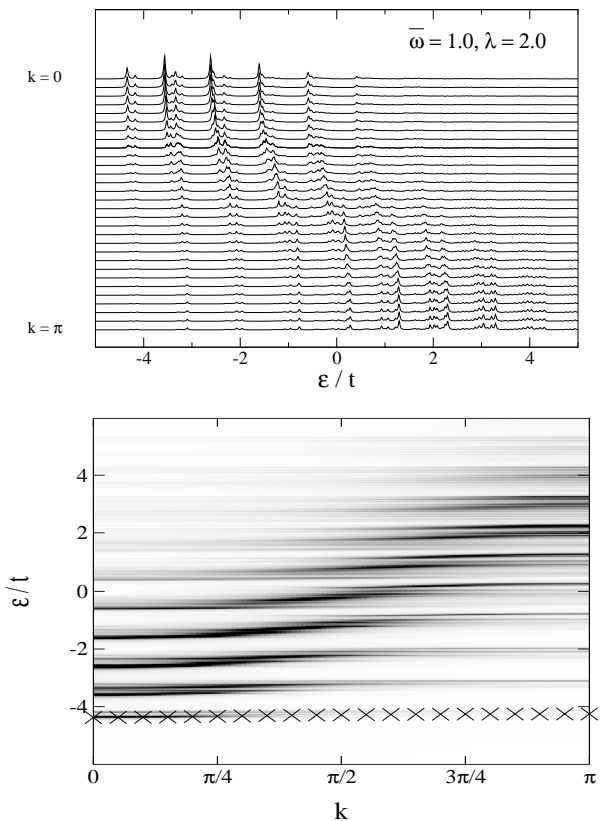}
\caption{\label{fig:spectrum1D_lambda2.0}
  Top: Spectral function $\sf$ of the 1D Holstein polaron calculated with CPT
  for $N=4$, $N_\text{ph}=25$ and $\eta=0.02t$. Bottom: Density plot of the
  same data for $100$ points in $k$ space. Symbols represent results of
  Bon\v{c}a.\cite{Bo03}}
\end{figure}

Finally, in Fig.~\ref{fig:spectrum1D_lambda2.0}, we report the spectral
function for $\omb=1$ and $\lambda=2.0$. The results have been obtained using
only a four-site cluster, which is sufficient to get very good agreement with
Bon\v{c}a's data for $E(k)$, with only minor deviations at large values of
$k$ where finite-size effects are most pronounced, as discussed in
Sec.~\ref{sec:comp-with-exact}. This is a consequence of the predominantly
local effects in the strong-coupling regime, which also manifest themselves
in terms of a very narrow polaron band.


\subsection{Results: Two dimensions}\label{sec:2d-holstein-polaron}

To illustrate the applicability of CPT, we also calculated the spectral
function of the Holstein polaron on a 2D cluster with $N=8$, which has the
shape of a tilted square. In contrast to CPT in one dimension, the choice for
the shape of the cluster is not unique, and different clusters may lead to
slightly different results. This possibility has been investigated for the
Hubbard model, and the effect of the cluster shape on $\sfb$ was found to be
rather small.\cite{SePePl02} For the case of the Holstein polaron, where the
physics is dominated by local correlations, the influence of the geometry of
the cluster is expected to be even smaller.

As discussed, \eg, by Wellein \etal,\cite{WeRoFe96} similar to 1D, a small
polaron is formed in two dimensions provided that $\lambda>1$ and
$\lambda/\omb>0.5$. While the behavior in the non-adiabatic regime ($\omb>1$)
is only weakly affected by dimensionality,\cite{WeRoFe96,CaCiGr98} important
differences exist in the adiabatic regime $\omb<1$: In contrast to the
one-dimensional case, where a large polaron is formed for any $\lambda>0$,
the electron remains quasi-free for $\lambda<1$, as indicated by an almost
unaffected effective hopping amplitude. Moreover, for the same value of
$\omb$, the cross-over to a small polaron at $\lambda\approx1$ is much
sharper in 2D than in 1D.

\begin{figure}
  \includegraphics[width=0.41\textwidth]{2d_sf.eps}\\\vspace*{1em}
  \includegraphics[width=0.45\textwidth,height=0.3\textwidth]{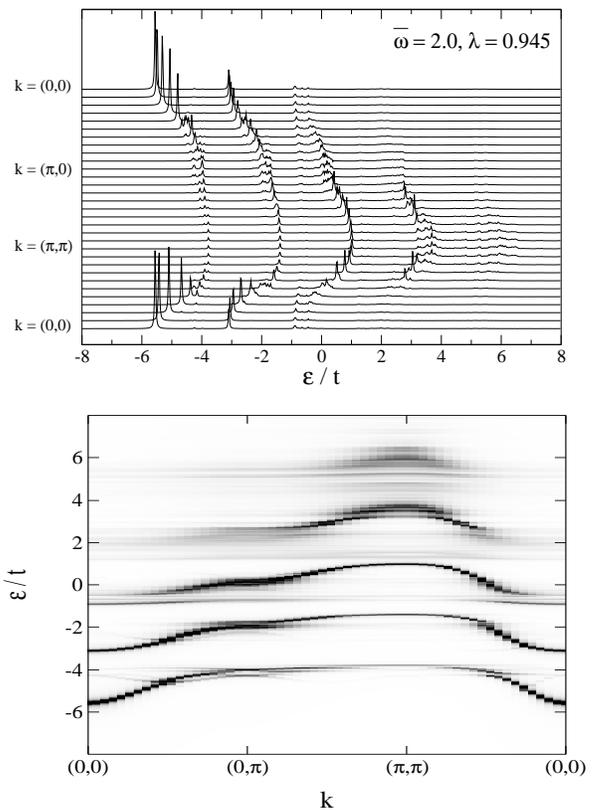}
\caption{\label{fig:spectrum2D}
  Top: Spectral function $\sfb$ of the 2D Holstein polaron calculated with
  CPT for $N=8$, $N_\text{ph}=9$ and $\eta=0.02t$. Bottom: Density plot of
  the same data (see text).}
\end{figure}

Here we simply aim to demonstrate the possibility of calculating the 2D
spectral function with CPT. Therefore, we restrict ourselves to one set of
parameters, namely $\omb=2.0$ and $\lambda=0.945$, which has also been
treated using {\it finite-cluster} strong-coupling perturbation
theory\cite{Stephan}. In contrast to standard SCPT based on the Lang-Firsov
transformation, the latter has been shown to give reliable results also for
intermediate $\lambda$ and $\omb$, which is a consequence of the inclusion of
longer-ranged effects.\cite{Stephan,WeFe97} While in the 1D case the density
plot of $\sfb$ contains all 100 values of $k$ used in CPT, in two dimensions
we have used 400 points in $\bk$ space. However only 60, lying along
$\Gamma\text{MX}\Gamma$, are shown in Fig.~\ref{fig:spectrum2D}.

From the above discussion, and for the parameters considered here, we expect
a rather broad polaron band. This is clearly confirmed by the spectral
function shown in Fig.~\ref{fig:spectrum2D}, and the lowest-energy band in
our data resembles closely the findings of Stephan (Fig.~2 of
Ref.~\onlinecite{Stephan}). In particular, similar to the one-dimensional
case considered in Sec.~\ref{sec:1d-holstein-polaron}, a flattening of the
polaron band near $(\pi/2,\pi)$ is found which has also been noted by Wellein
\etal\cite{WeRoFe96} Above the polaron band, also similar to 1D, there lie
several other incoherent bands which correspond to multi-phonon excitations
and are therefore separated in energy by $\omb$.

In summary, the results of this section clearly demonstrate that CPT is
applicable not only in the strong-coupling regime, but also for weak and
intermediate electron-phonon interaction. The quality of the resulting
spectra is superior to ED data for the same cluster size, and very good
agreement has been found with the variational method of Bon\v{c}a
\etal\cite{BoTrBa99} in one dimension. Moreover, we have shown that CPT also
allows accurate calculations of $\sfb$ in two dimensions.

\section{\label{sec:holsteinDE}Many-electron case}

In the last section, we have restricted ourselves to the Holstein model with
one electron. Although CPT has been successfully applied, \eg, to the
many-electron Hubbard model,\cite{SePePL00,SePePl02,DaArHa02,AiDaEvvdL02} the
electron-phonon coupling in the Holstein model greatly complicates
calculations using Lanczos ED.  For finite electron density, we combine CPT
with the exact analytic result for the Green function in the atomic limit.
For the atomic limit, the Green function has been obtained for many models
using the equation-of-motion method,\cite{Ma90} and here it will allow us to
obtain results for the many-electron case which will be compared with the
many-body coherent potential approximation discussed below.

\subsection{Many-body coherent potential approximation}\label{sec:many-body-cpa}

Extending previous work of Edwards \etal\cite{EdGrKu99,GrEd99} for the pure
double-exchange (DE) model (see, \eg, Ref.~\onlinecite{David_AiP}),
Green\cite{Gr01} studied the Holstein-DE model using a many-body coherent
potential approximation (CPA) which, owing to the more complicated form of
the Holstein-DE Hamiltonian, constitutes a considerable extension of the
Hubbard III approximation.\cite{Hu64} The many-body CPA successfully
describes many aspects of the manganites, and we refer the reader to a recent
review of this work by Edwards.\cite{David_AiP} Here we only consider the
special case of a completely saturated ferromagnetic state at temperature
$T=0$, with all itinerant spins having $\up$ spin, say. Consequently, the DE
term which couples local and itinerant spins\cite{David_AiP} becomes merely a
constant shift in energy, and the Holstein-DE model is equivalent to the pure
Holstein model of spinless fermions, \ie with no doubly-occupied
sites.\cite{HoEd01} An important feature of the many-body CPA is that the
one-electron Green function reduces to the exact atomic limit for $t=0$,
which takes the form\cite{HoEd01}
\begin{equation}\label{eq:G_AL}
G^\text{AL}_\up(\en) = \rme^{-\alpha}
\left\{
\frac{1}{\en} + \sum_{r=1}^\infty\frac{\alpha^r}{r!}
\left(\frac{n}{\en+\om r}+\frac{1-n}{\en-\om r}\right)
\right\}\,,
\end{equation}
where $\alpha=g^2/\om^2$ and the polaron binding energy $-(g^2/\om)n$ ($n=0$,
1) has been absorbed into the chemical potential.  The general result for
$G^\text{AL}$ of the Holstein model with electrons of both spins has been
given by Green,\cite{Gr01} and we drop the spin-index in the sequel.  As
discussed by Edwards,\cite{David_AiP} for an elliptic density of states, the
local Green function $\G(z)$ for complex energy $z$ satisfies the CPA
equation
\begin{equation}\label{eq:CPA}
\G(z) = G^\text{AL}(z-W^2\G/4)
\end{equation}
and the self-energy can be obtained from\cite{David_AiP}
\begin{equation}\label{eq:Sigma}
\Sigma(z) = z - \G^{-1} - W^2 \G /4\,.
\end{equation}
Finally, the one-electron spectral function is given by
\begin{equation}\label{eq:sf}
\sfb = -\pi^{-1}\Im\,[z - \ek - \Sigma(z)]\,,
\end{equation}
where
\begin{equation}\label{eq:band}
  \ek=-2t\sum_{m=1}^3\cos k_m
\end{equation}
is the band energy for wavevector $\bk$.

\begin{figure}
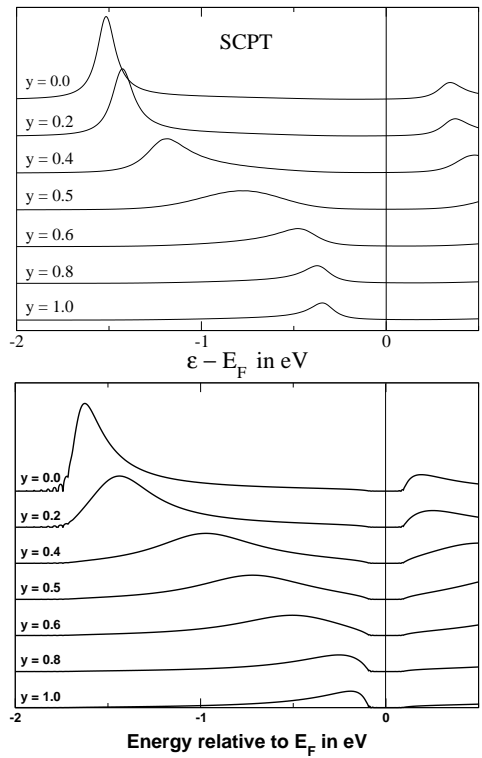

  \includegraphics[width=0.35\textwidth]{SCPT.eps}\\%
  \includegraphics[width=0.35\textwidth]{CPA.eps}
\caption{\label{fig:CPA-CPT}
  Comparison of the spectral function of the Holstein model of spinless
  fermions at $T=0$, calculated with SCPT (top) and with the many-body CPA
  (bottom, taken from Ref.~\onlinecite{HoEd01}). Here the wavevector $\bk$ is
  given by $\bk=\pi(1,y,0)$ with $y$ as indicated in the figure. The plot is
  for $\om/W=0.05$ and $g/W=0.2$. The SCPT results have been broadened using
  a smearing parameter $\eta/W=0.05$.}
\end{figure}

In order to compare with angle-resolved photoemission (ARPES) data on the
bilayer manganite \chem{La_{1.2}Sr_{1.8}Mn_2O_7}, nominally with $n=0.6$,
Hohenadler and Edwards chose a strong-electron phonon coupling $g/W=0.2$, as
deduced from the low Curie temperature of this material.\cite{HoEd01} To
simplify calculations, they also used $n=0.5$ for which case the chemical
potential $\mu=0$ by symmetry. We want to point out that the many-body CPA
assumes a homogeneous system, so that no tendencies toward
charge-density-wave order occur as $n$ is varied.\cite{Gr01} As in previous
work,\cite{Gr01} Hohenadler and Edwards used $W=1$eV and $\om/W=0.05$ (see
also Ref.~\onlinecite{David_AiP}). The results\cite{HoEd01} for $\sfb$, shown
in Fig.~\ref{fig:CPA-CPT}, support the theory of Alexandrov and
Bratkovsky\cite{AlBrat99} that in these manganites, small polarons exist in
the ferromagnetic state. A similar interpretation of the experimental
data--based on standard small-polaron theory--had also been given by Dessau
et al.\cite{De98} Well away from the Fermi surface, a well-defined peak
exists which broadens as $\bk$ approaches the Fermi level $E_\text{F}$ at
$y=0.5$. If $y$ is increased further, most of the spectral weight is
transfered above $E_\text{F}$.  Moreover, the peaks never approach the Fermi
level closely, in agreement with the experimental data. This indicates the
existence of a pseudogap in the one-electron density of states.  However, in
the gap, there exist small polaron sub-bands (see Fig.~4 of
Ref.~\onlinecite{Gr01}) and one of them, at the Fermi level, presumably gives
rise to the low but finite conductivity of the system.  As discussed by
Edwards,\cite{David_AiP} the many-body CPA does not give coherent states with
infinite lifetime at the Fermi level, even for $T=0$.  This is typical for
any CPA, and here it leads to an incoherent polaron sub-band around the Fermi
level. Nevertheless, outside the central band around $E_\text{F}$, the
imaginary part of the self-energy displays the correct behavior, \ie it
vanishes in the gap, between the polaron bands.

\subsection{SCPT}\label{sec:scpt}

In this section we use the exact result for the atomic-limit Green function
of the Holstein model of spinless fermions, $G^\text{AL}$
(Eq.~(\ref{eq:G_AL})), and combine it with CPT to compare the resulting
spectrum with the many-body CPA. For this case of a single-site cluster
($a\equiv b$), Eqs.~(\ref{eq:CPT1}) and~(\ref{eq:CPT2}) reduce to a single
equation for the one-electron Green function\cite{SePePl02}
\begin{equation}\label{eq:GF_k}
\G(\bk,z) = \frac{G^\text{AL}(z)}{1-\ek G^\text{AL}(z)}
              = \frac{1}{z-\ek-\Sigma^\text{AL}(z)}
\end{equation}
with $z=\en+\rmi\eta$ and $\ek$ as defined by Eq.~(\ref{eq:band}).  Hence, as
mentioned before, CPT for $N=1$ is equivalent to the Hubbard I
approximation,\cite{Hu63} but here with the more complicated atomic-limit
Green function of the Holstein model given by Eq.~(\ref{eq:G_AL}). In the
sequel, we shall refer to this approximation as SCPT. This is justified by
the fact that the approach becomes exact for $t=0$. Historically, a similar
strong-coupling expansion for the Hubbard
model\cite{PaSeTr98,PaSeTr00}--including higher order corrections--has been
the starting point for the development of CPT.

Before we discuss the results, we would like to comment on the quality of the
SCPT used here: While the many-body CPA requires a self-consistent, iterative
solution of Eq.~(\ref{eq:CPA}), the SCPT Green function is obtained from the
Lehmann representation of the atomic-limit Green function~(\ref{eq:G_AL}),
and the subsequent use of the resulting self-energy $\Sigma^\text{AL}$ in
Eq.~(\ref{eq:GF_k}). Similar to the original Hubbard I
approximation,\cite{Hu63} the resulting Green function consists of delta
peaks corresponding to states with infinite lifetime. However, due to the
poles in the self-energy, there are no states at the Fermi level and the
system is not a Fermi liquid.  As in the many-body CPA, $\G$ depends on $\bk$
only through the band energy $\ek$, whereas the self-energy is local. This
reliance on the atomic limit is reasonable in the strong-coupling regime
considered here, where small polarons move in an extremely narrow band.
Consequently, the simple perturbative treatment of the hopping term can be
expected to give sensible results. Nevertheless, in SCPT, we have to use an
artificial imaginary part $\eta$--which does not depend on energy--to obtain
peaks of finite width.  Although for large enough $\eta$ there will be states
at the Fermi level, the latter have only finite lifetime even for $T=0$.
Hence, both SCPT and the many-body CPA never give a Fermi liquid, but the
self-consistent CPA Green function yields an imaginary part of the
self-energy that shows the correct, strong energy-dependence except for the
region inside the very small, incoherent polaron band around $E_\text{F}$, as
discussed in Sec.~\ref{sec:many-body-cpa}. Thus, as could be expected from
the Hubbard I-like approximation in Eq.~(\ref{eq:GF_k}), the many-body CPA is
superior to SCPT, although both approaches become exact in the atomic limit.

The spectral function obtained with SCPT using Eq.~(\ref{eq:Akw}), also shown
in Fig.~\ref{fig:CPA-CPT}, resembles quite closely the results of Hohenadler
and Edwards.\cite{HoEd01} Although there are some differences concerning the
width and the position of the peaks, the overall behavior is very similar. In
particular, the broadening of the QP peak near the Fermi surface at $y=0.5$
is well reproduced.  Clearly, the success of SCPT consists in a surprisingly
good agreement with the CPA data for all $\bk$.  Despite this agreement, CPT
fails to reproduce the polaron sub-bands, and the sharp edge to the pseudogap
for large values of $y$. Moreover, the gap is larger than in the CPA data.
These shortcomings are a consequence of the rather crude approximation.
Nevertheless, keeping in mind the simplicity of the ansatz, the agreement
with the many-body CPA is satisfactory.  We would like to point out that the
SCPT presented here can also be generalized to the Holstein-DE model with
quantum spins (\eg $S=3/2$ appropriate for the manganites\cite{David_AiP})
and at finite temperature, using the atomic-limit Green function given by
Edwards.\cite{David_AiP} Finally, the approximation could be systematically
improved by increasing the number of sites in the cluster, which is exactly
the idea behind CPT.

However, for $N>1$ the cluster Green function can no longer be calculated
analytically and one has to resort to numerical methods such as ED as in
Sec.~\ref{sec:holsteinpolaron}. Such calculations are extremely difficult for
the case of quarter-filled two- or three dimensional clusters, small phonon
frequency and strong electron-phonon coupling.  Future work along these
lines--employing optimized phonon approaches\cite{WeFeWeBi00} (see also
Sec.~\ref{sec:summary})--is highly desirable in order to assess the quality
of the many-body CPA results.

\section{\label{sec:summary}Conclusions}

We have applied cluster perturbation theory to the Holstein polaron problem
in one and two dimensions, and comparison with existing work has revealed
very good agreement. In combination with the Lanczos method to calculate the
cluster Green function, the method gives reliable results for the
one-electron spectral function $\sfb$, which become exact in the weak- and
strong-coupling limit, $\lambda=0$ and $t=0$, respectively, and for the case
of an infinite cluster. Calculations for continuous values of the wavevector
$\bk$ are possible and, more importantly, finite-size effects are
significantly reduced compared to standard ED. Our results extend previous
applications of CPT to Hubbard and $t-J$ models, showing that the method is
also well-suited for electron-phonon models with local interactions.  As
pointed out before, for more than one electron in the system, it becomes
increasingly difficult to include enough phonon states so as to obtain
converged results. Future work may therefore combine optimized phonon
approaches\cite{WeFeWeBi00} and CPT to investigate more complicated problems
such as the many-electron case, or extended models including, \eg, a Hubbard
term. The major advantage of such an approach is the possibility of using
relatively small clusters while still obtaining results which are only weakly
influenced by finite-size effects.

Additionally, we have used the exact atomic-limit Green function of the
Holstein model of spinless fermions, to calculate the spectral function for
the case quarter-filling and strong electron-phonon coupling. The results of
this approximation at the Hubbard I level agree surprisingly well with the
many-body CPA.

\begin{acknowledgments}
  
  M.H. and M.A. were supported by DOC [Doctoral Scholarship Program of the
  Austrian Academy of Sciences]. We are grateful to Janez Bon\v{c}a for
  providing us with previously unpublished data, and we acknowledge helpful
  discussions with David S\'{e}n\'{e}chal and Maria Daghofer. Finally, we
  would like to thank David M. Edwards for making valuable comments on the
  manuscript.

\end{acknowledgments}



\end{document}